\documentclass[11pt,letterpaper]{article}
\usepackage{graphicx}% Include figure files
\usepackage{dcolumn}% Align table columns on decimal point
\usepackage{bm, amsmath, amssymb}% bold math
\usepackage{subfig}
\usepackage{graphicx,epic,eepic,epsfig,amsmath,latexsym,amssymb,verbatim,color}
\usepackage{theorem}

\usepackage{calc}

\newlength{\myrightmargin}
\newlength{\myleftmargin}
\newlength{\mytopmargin}
\newlength{\mybottommargin}

\def\squareforqed{\hbox{\rlap{$\sqcap$}$\sqcup$}}
\def\qed{\ifmmode\squareforqed\else{\unskip\nobreak\hfil
\penalty50\hskip1em\null\nobreak\hfil\squareforqed
\parfillskip=0pt\finalhyphendemerits=0\endgraf}\fi}
\def\endenv{\ifmmode\;\else{\unskip\nobreak\hfil
\penalty50\hskip1em\null\nobreak\hfil\;
\parfillskip=0pt\finalhyphendemerits=0\endgraf}\fi}

\mathchardef\ordinarycolon\mathcode`\:
\mathcode`\:=\string"8000
\def\vcentcolon{\mathrel{\mathop\ordinarycolon}}
\begingroup \catcode`\:=\active
  \lowercase{\endgroup
  \let :\vcentcolon
  }

\newcommand{\nc}{\newcommand}
\nc{\rnc}{\renewcommand}
\nc{\beq}{\begin{equation}}
\nc{\eeq}{{\end{equation}}}
\nc{\beqa}{\begin{eqnarray}}
\nc{\eeqa}{\end{eqnarray}}
\nc{\lbar}[1]{\overline{#1}}
\nc{\bra}[1]{\langle#1|}
\nc{\ket}[1]{|#1\rangle}
\nc{\ketbra}[2]{|#1\rangle\!\langle#2|}
\nc{\braket}[2]{\langle#1|#2\rangle}
\nc{\proj}[1]{| #1\rangle\!\langle #1 |}
\nc{\avg}[1]{\langle#1\rangle}
\nc{\Rank}{\operatorname{Rank}}
\nc{\smfrac}[2]{\mbox{$\frac{#1}{#2}$}}
\nc{\Tr}{\operatorname{Tr}}
\nc{\tr}{\operatorname{Tr}}
\nc{\1}{\openone}
\nc{\ox}{\otimes}
\nc{\dg}{\dagger}
\nc{\dn}{\downarrow}
\nc{\cA}{{\cal A}}
\nc{\cB}{{\cal B}}
\nc{\cC}{{\cal C}}
\nc{\cD}{{\cal D}}
\nc{\cE}{{\cal E}}
\nc{\cF}{{\cal F}}
\nc{\cG}{{\cal G}}
\nc{\cH}{{\cal H}}
\nc{\cI}{{\cal I}}
\nc{\cJ}{{\cal J}}
\nc{\cK}{{\cal K}}
\nc{\cL}{{\cal L}}
\nc{\cM}{{\cal M}}
\nc{\cN}{{\cal N}}
\nc{\cO}{{\cal O}}
\nc{\cP}{{\cal P}}
\nc{\cQ}{{\cal Q}}
\nc{\cR}{{\cal R}}
\nc{\cS}{{\cal S}}
\nc{\cT}{{\cal T}}
\nc{\cX}{{\cal X}}
\nc{\cY}{{\cal Y}}
\nc{\cZ}{{\cal Z}}
\nc{\supp}{{\operatorname{supp}}}
\nc{\var}{\operatorname{var}}
\nc{\rar}{\rightarrow}
\nc{\lrar}{\longrightarrow}
\nc{\polylog}{\operatorname{polylog}}

\nc{\RR}{{{\mathbb R}}}
\nc{\FF}{{{\mathbb F}}}
\nc{\NN}{{{\mathbb N}}}
\nc{\ZZ}{{{\mathbb Z}}}
\nc{\PP}{{{\mathbb P}}}
\nc{\QQ}{{{\mathbb Q}}}
\nc{\UU}{{{\mathbb U}}}
\nc{\EE}{{{\mathbb E}}}
\nc{\Icoh}{{I^{\rm coh}}}
\nc{\Qca}{{Q_{\rm ss}}}
\nc{\Qcaa}{{Q^{(1)}_{\rm ss}}}
\nc{\Dcaa}{{D^{(1)}_{{\rm ss}\rightarrow}}}
\nc{\Dca}{{D_{{\rm ss}\rightarrow}}}

\nc{\be}{\begin{equation}}
\nc{\ee}{{\end{equation}}}
\nc{\bea}{\begin{eqnarray}}
\nc{\eea}{\end{eqnarray}}
\nc{\<}{\langle}
\rnc{\>}{\rangle}
\nc{\Hom}[2]{\mbox{Hom}(\CC^{#1},\CC^{#2})}
\nc{\rU}{\mbox{U}}

\def\pmat#1{{\begin{pmatrix}#1\end{pmatrix}}}

\nc{\CA}{{\cal A}} \nc{\CB}{{\cal B}} \nc{\CC}{{\cal C}}
\nc{\CD}{{\cal D}} \nc{\CE}{{\cal E}} \nc{\CF}{{\cal F}}
\nc{\CG}{{\cal G}} \nc{\CH}{{\cal H}} \nc{\CI}{{\cal I}}
\nc{\CJ}{{\cal J}} \nc{\CK}{{\cal K}} \nc{\CL}{{\cal L}}
\nc{\CM}{{\cal M}} \nc{\CN}{{\cal N}} \nc{\CO}{{\cal O}}
\nc{\CP}{{\cal P}} \nc{\CQ}{{\cal Q}} \nc{\CR}{{\cal R}} \nc{\CS}{{\cal S}}
\nc{\CT}{{\cal T}} \nc{\CU}{{\cal U}} \nc{\CV}{{\cal V}}
\nc{\CX}{{\cal X}} \nc{\CW}{{\cal W}} \nc{\CZ}{{\cal Z}}

\nc{\bA}{\mathbb{A}} \nc{\bB}{\mathbb{B}} \nc{\bC}{\mathbb{C}}
\nc{\bD}{\mathbb{D}} \nc{\bE}{\mathbb{E}} \nc{\bF}{\mathbb{F}}
\nc{\bG}{\mathbb{G}} \nc{\bH}{\mathbb{H}} \nc{\bI}{\mathbb{I}}
\nc{\bJ}{\mathbb{J}} \nc{\bK}{\mathbb{K}} \nc{\bL}{\mathbb{L}}
\nc{\bM}{\mathbb{M}} \nc{\bN}{\mathbb{N}} \nc{\bO}{\mathbb{O}}
\nc{\bP}{\mathbb{P}} \nc{\bQ}{\mathbb{Q}} \nc{\bR}{\mathbb{R}} \nc{\bS}{\mathbb{S}}
\nc{\bT}{\mathbb{T}} \nc{\bU}{\mathbb{U}} \nc{\bV}{\mathbb{V}}
\nc{\bX}{\mathbb{X}} \nc{\bW}{\mathbb{W}} \nc{\bZ}{\mathbb{Z}}

\nc{\msA}{\mathscr{A}} \nc{\msB}{\mathscr{B}} \nc{\msC}{\mathscr{C}}
\nc{\msD}{\mathscr{D}} \nc{\msE}{\mathscr{E}} \nc{\msF}{\mathscr{F}}
\nc{\msG}{\mathscr{G}} \nc{\msH}{\mathscr{H}} \nc{\msI}{\mathscr{I}}
\nc{\msJ}{\mathscr{J}} \nc{\msK}{\mathscr{K}} \nc{\msL}{\mathscr{L}}
\nc{\msM}{\mathscr{M}} \nc{\msN}{\mathscr{N}} \nc{\msO}{\mathscr{O}}
\nc{\msP}{\mathscr{P}} \nc{\msQ}{\mathscr{Q}} \nc{\msR}{\mathscr{R}} 
\nc{\msS}{\mathscr{S}} \nc{\msT}{\mathscr{T}} \nc{\msU}{\mathscr{U}} 
\nc{\msV}{\mathscr{V}} \nc{\msX}{\mathscr{X}} \nc{\msW}{\mathscr{W}} 
\nc{\msY}{\mathscr{Y}} \nc{\msZ}{\mathscr{Z}}

\nc{\mfa}{{\mathfrak a}} \nc{\mfb}{{\mathfrak b}} \nc{\mfc}{{\mathfrak c}}
\nc{\mfd}{{\mathfrak d}} \nc{\mfe}{{\mathfrak e}} \nc{\mff}{{\mathfrak f}}
\nc{\mfg}{{\mathfrak g}} \nc{\mfh}{{\mathfrak h}} \nc{\mfi}{{\mathfrak i}}
\nc{\mfj}{{\mathfrak j}} \nc{\mfk}{{\mathfrak k}} \nc{\mfl}{{\mathfrak l}}
\nc{\mfm}{{\mathfrak m}} \nc{\mfn}{{\mathfrak n}} \nc{\mfo}{{\mathfrak o}}
\nc{\mfp}{{\mathfrak p}} \nc{\mfq}{{\mathfrak q}} \nc{\mfr}{{\mathfrak r}}
\nc{\mfs}{{\mathfrak s}} \nc{\mft}{{\mathfrak t}} \nc{\mfu}{{\mathfrak u}}
\nc{\mfv}{{\mathfrak v}} \nc{\mfw}{{\mathfrak w}} \nc{\mfx}{{\mathfrak x}}
\nc{\mfy}{{\mathfrak y}} \nc{\mfz}{{\mathfrak z}}

\nc{\mfA}{{\mathfrak A}} \nc{\mfB}{{\mathfrak B}} \nc{\mfC}{{\mathfrak C}}
\nc{\mfD}{{\mathfrak D}} \nc{\mfE}{{\mathfrak E}} \nc{\mfF}{{\mathfrak F}}
\nc{\mfG}{{\mathfrak G}} \nc{\mfH}{{\mathfrak H}} \nc{\mfI}{{\mathfrak I}}
\nc{\mfJ}{{\mathfrak J}} \nc{\mfK}{{\mathfrak K}} \nc{\mfL}{{\mathfrak L}}
\nc{\mfM}{{\mathfrak M}} \nc{\mfN}{{\mathfrak N}} \nc{\mfO}{{\mathfrak O}}
\nc{\mfP}{{\mathfrak P}} \nc{\mfQ}{{\mathfrak Q}} \nc{\mfR}{{\mathfrak R}}
\nc{\mfS}{{\mathfrak S}} \nc{\mfT}{{\mathfrak T}} \nc{\mfU}{{\mathfrak U}}
\nc{\mfV}{{\mathfrak V}} \nc{\mfW}{{\mathfrak W}} \nc{\mfX}{{\mathfrak X}}
\nc{\mfY}{{\mathfrak Y}} \nc{\mfZ}{{\mathfrak Z}}

\def\id{{\mathbb I}}
\begin{document}

\title{Gaussian bosonic synergy:\\ quantum communication via realistic channels of
zero quantum capacity }% Force line breaks with \\

\author{Graeme Smith$^1$, John A. Smolin$^1$ and Jon Yard$^2$ \\ \\
{\it \small $^1$IBM T.J. Watson Research Center, Yorktown Heights, NY 10598, USA }\\
{\it \small $^2$Computational and Computer Sciences (CCS-3) and Center for Nonlinear Studies (CNLS) }\\  
{\it \small Los Alamos National Laboratory, Los Alamos, NM 87545, USA }
}

\date{\today}% It is always \today, today,
             % but any date may be explicitly specified

\maketitle
\renewcommand{\abstractname}{}
\begin{abstract}
\vspace{-.4in} { As with classical information, error-correcting codes enable reliable transmission of quantum information through noisy or lossy channels. In contrast to the classical theory, imperfect quantum channels exhibit a strong kind of synergy: there exist pairs of discrete memoryless quantum channels, each of zero quantum capacity, which acquire positive quantum capacity when used together.  Here we show that this ``superactivation" phenomenon also occurs in the more realistic setting of optical channels with attenuation and Gaussian noise.  This paves the way for its experimental realization and application in real-world communications systems.
}
\end{abstract}

In the spirit of Shannon's information theory \cite{Shannon48}, any
physical process acting on a quantum system can be
thought of as a communication channel.  One may then speak of its
capacity for transmitting quantum states as the fundamental amount of quantum information that can be protected using error
correction.  Quantum capacity measures the number of qubits, or
two-level quantum systems, that can be protected with vanishing error
in the limit of many channel uses.  In contrast to Shannon's capacity, no simple formula is
known for the quantum capacity.  While finite-dimensional systems like qubits are convenient
units for quantifying quantum information and for describing abstract
protocols, real-world applications require consideration of
continuous-variable systems, such as optical, electromagnetic, and more
general bosonic systems.

Photons are the natural carriers of information in radio, cellular,
fiber and free-space optical networks.  Since bosons mediate
the fundamental forces of nature, understanding information flows in bosonic systems
is of both deep theoretical and practical importance.  Noise in 
conventional networks is often well-approximated
by additive gaussian noise, and the six decades since Shannon's
theory was introduced have seen the emergence of a mature
theory of communication in such practical networks.  Low power optical
noise, on the other hand, is quantum mechanical in nature.  Despite
considerable recent progress,
comparatively simple questions about point-to-point capacities of
optical channels with gaussian noise, or even finite-dimensional
channels, remain unanswered.  Nonetheless, commercial quantum
networks are being deployed worldwide for fundamentally quantum tasks
like quantum cryptography \cite{BB84}, while experimental development
of quantum memories paves the way toward the quantum repeaters and
quantum computers of the future.

Gaussian quantum noise is a generalization of the additive white
gaussian noise at the heart of classical information theory
\cite{CoverThomas}.  It arises when optical modes unitarily interact
via a quadratic Hamiltonian with vacuum environment modes \cite{HW01,EW05}.  Alternatively, a gaussian
state is a state with a gaussian characteristic function and a gaussian channel maps
gaussian states to gaussian states.  Examples of gaussian states
include thermal, coherent, and squeezed states.  Fock, or number
states, are not gaussian and are much more difficult to
produce.  Table 1 describes gaussian states and channels in detail.

\begin{table}[ht]
\centering  % used for centering table
\begin{tabular}{|c |c |c|} % centered columns (4 columns)
\hline\hline                        %inserts double horizontal lines
  & General State & Gaussian State\\
\hline                  % inserts single horizontal line
& & \\
 & $\Tr \rho e^{\mathbf{v}^tJ\mathbf{R}}$ & $e^{\mathbf{v}^t \mathbf{d} - \frac{1}{4}\mathbf{v}^t \gamma\mathbf{v}}$\\ 
Characteristic& & \\
Function & where $\mathbf{R} =(Q_1\,\, P_1\, \,\,\cdots\,\, Q_m \,\,P_m)^t$ &  with $d_i = \langle R_i\rangle$ and \\
& & $\gamma_{ij} = \langle R_iR_j+R_jR_i\rangle-d_id_j$\\
& & \\
\hline %inserts single line
& & \\
& $[R_j,R_k] = i J_{jk}$& \\
Uncertainty & & $\gamma + i J \geq 0$\\
Relations &  where $J =\left(\!\!\begin{smallmatrix}\,\,\,\,0 & 1 \\ -1 &0\end{smallmatrix}\right)^{\oplus n}$ & \\
& & \\
\hline
 & & \\
  & $\rho \rightarrow \Tr_E U \rho \otimes \proj{0} U^\dagger$ & $\gamma \rightarrow X \gamma X^t + Y$\\
Noisy & & \\
Evolution& $U$ unitary &  $Y = Y^t$ and \\
	& & $Y + i (J - X J X^t) \geq 0$\\
 & & \\
\hline
\end{tabular}
\caption{\small General vs Gaussian bosonic states.  The quantum phase space $A$ of $m$ bosonic
modes is described by canonical coordinates $q_1,p_1,\dotsc,q_m,p_m$.
A corresponding vector of canonical operators $\mathbf{R}$ acts on the underlying infinite-dimensional
quantum Hilbert space $\CH_A$ and satisfies Heisenberg's uncertainty
relations $[R_j,R_k] = i J_{jk}$.  A state of $A$ is gaussian
precisely when its characteristic function is a gaussian function of the phase space vector
${\mathbf{v}}$. Such a state is is characterized by its displacement
vector $\mathbf{d}$ and covariance matrix $\gamma$, both defined in terms of expectation values 
$\langle R_i\rangle = \Tr(R_i\rho)$ and $\langle R_iR_j\rangle = \Tr(R_iR_j\rho)$. 
Some examples of covariance matrices are the covariance of the vacuum, $\id_2$,  a single-mode
thermal (Bose-Einstein) state with average photon number $n$ which has
covariance matrix $(2n + 1)\id_2$, as well as a squeezed  vacuum state with $\gamma = {\rm diag}(\eta, 1/\eta)$.    Disregarding phase space displacements, gaussian channel can be described by a linear map from covariance
matrices on a set of input modes $A$, to a set of output modes $B$.}
 \label{table:GaussianDefs} % is used to refer this table in the text
\end{table}

Any noisy quantum evolution $\CN\colon A\to B$ from states on $\CH_A$
to states on $\CH_B$ can be mathematically extended to a unitary
interaction of an input state with an uncorrelated and inaccessible
environment with Hilbert space $\CH_E$.  This defines a second,
complementary channel to the environment  $\CH_E$, that appears in a useful lower bound to the quantum
capacity.  This lower bound is the maximum of the \emph{coherent
  information}, or difference $H(B) - H(E)$ between the entropies of
the output and environment, maximized over all input states on
$\CH_A$, where the von Neumann entropy of a density matrix
$\rho$ is $H= -\tr \rho \log_2 \rho$.  It is a lower bound in the
sense that good sequences of error-correcting codes achieving this
rate exist \cite{D03,Shor02,Lloyd97}.  Only in special cases can we
effectively calculate the optimization implicit in the coherent
information, let alone the quantum capacity itself. 

Quantum capacities of gaussian channels have been considered in detail \cite{HW01,HP01}, where coherent information was calculated to
give a lower bound for several examples.  The quantum capacity of the lossy bosonic channel was found in \cite{WGC06}.  Classical
and cryptographic capacities of bosonic gaussian channels have also
been considered by many authors
\cite{HW01,HP01,GGLMSY04,WGC06,WPG07}. However, even restricting to
gaussian noise does not appear to make the problem solvable
and the potential exists for exotic behavior for these channels.

Pairs of zero quantum capacity channels can nevertheless allow
noiseless communication when used together \cite{SY08}.  
This {\em superactivation}  arises when two zero-capacity channels have
very different noise properties.  The weakness of each channel is
overcome by the strengths of the other, illustrating that the
communication capability of a channel is not a simple function of the
channel alone, but also depends on the context in which it is used.
The discovery of \cite{SY08} followed a sequence of superadditivity
findings in the multi-party and two-way settings
\cite{H97,HHH99,SST00,NatPhys4}, and was followed by several
substantial discoveries in the conventional one-way setting
\cite{Hastings09,SS09,LWZG09,BO2010,CKCH10}.  Below we present some simple and natural examples of superactivation with
gaussian channels that can potentially be realized with current
technologies, demonstrating the richness of the set of gaussian
channels and the complexity of their capacity-achieving protocols.
Superactivation is therefore not merely an oddity confined to
unrealistic models but is in fact {\em necessary} for a proper
characterization of realistic communication settings.

\begin{figure}
\centering
\includegraphics[width=4in]{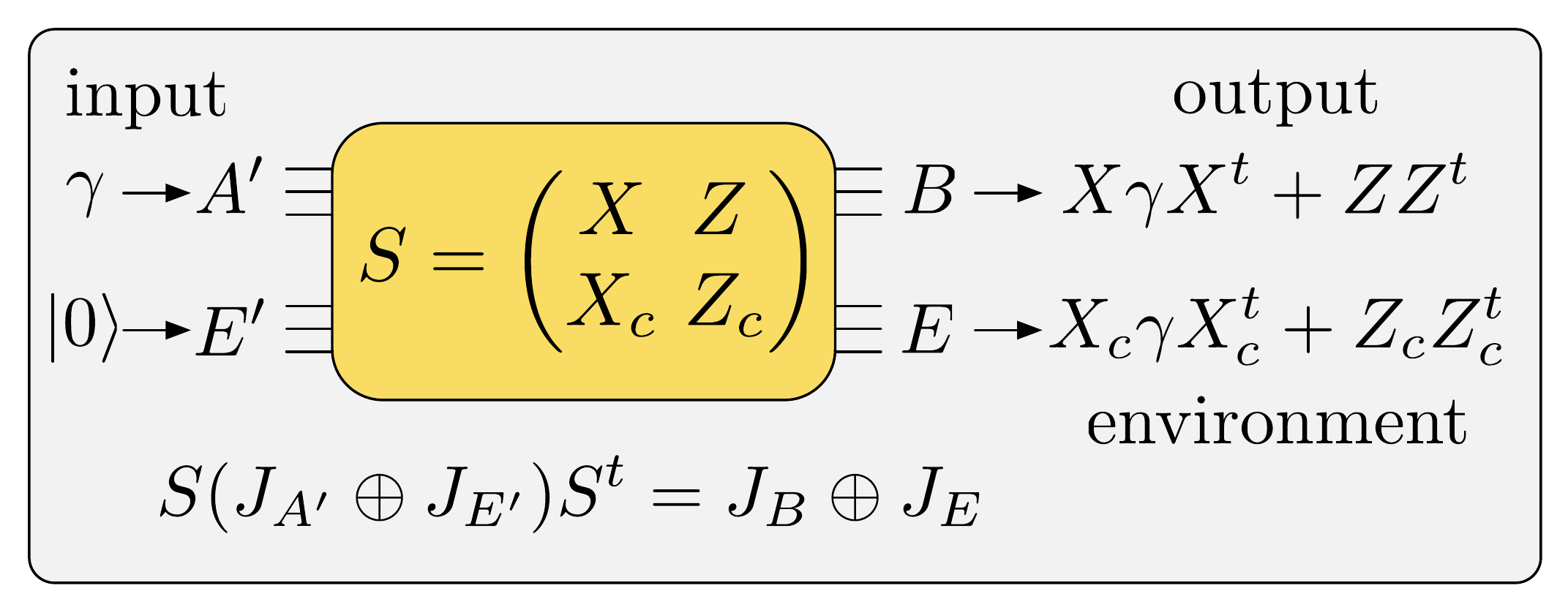}
\caption{\small Just as an arbitrary quantum channel can be implemented as a unitary transformation acting on a larger space, every gaussian quantum channel can be represented by a symplectic matrix $S$ as illustrated as a block matrix in the figure.  The corresponding channel maps
%$S = \left(\begin{smallmatrix}X & Z \\ X_c & Z_c\end{smallmatrix}\right)$, 
the input $A'$ to the output $B$ while the complement maps it to the
environment $E$.  The ancillary input modes $E'$ are assumed to be in
the vacuum state with covariance $\id_{E'}$.  The matrix $S$ is
required to be symplectic, or canonical transformation, meaning that
it satisfies the equation at the bottom of the figure, and is thus
compatible with the symplectic structures of the input and output
modes.  }
\label{fig:symplectic}
\end{figure}

There are two classes of channels known to have zero quantum capacity. 
The first is the antidegradable channels \cite{DS03}, where the environment can simulate the output.
A simple example is a 50\% attenuation
channel, which is modeled by a beamsplitter. 
The other zero-capacity class is the entanglement binding channels
\cite{HHH98}, which produce a weak type of entanglement
between sender and receiver  that prohibits quantum communication and is analogous to thermodynamical bound
energy.  These PPT channels only produce states satisfying the 
positive partial transpose (PPT) nondistillability criterion \cite{Peres96,HHH96} which, 
for covariance matrix $\gamma_{AB}$ is $\gamma_{AB}+ i(J_A \oplus-J_B) \geq 0$ \cite{WW01}.
These are precisely the channels that remain physical when composed with the potentially
nonphysical operation of time-reversal.  A gaussian channel of the
form $\gamma \rightarrow X \gamma X^t + Y$ is PPT if and only if $Y + i (J + XJX^t)
\geq 0$.  

Since PPT bound entangled gaussian states can only occur when each
side has at least two modes \cite{WW01}, the smallest PPT channel one
might hope to superactivate would act on two modes.  We have found a
family of such examples, the simplest being
\begin{equation}
X = \pmat{-1 & 0 & 0 & 0 \\ 0 & 0 & 0 & 1 \\ 0 & 0 & 1 & 0 \\ 0 & 1 & 0 & 0},\,\,\,\,\,\,\,\,
 Y = \pmat{\sqrt{2} & 0 & 0 & 0\\  0 & \sqrt{2}& 0 & 0\\0 & 0 & \sqrt{2} & 0\\0 & 0 & 0 &\sqrt{2}}.
 \label{eqn:basicexample}
 \end{equation}
 Direct calculation reveals that each of the matrices $Y + i (J \pm X
 J X^t)$ has eigenvalues $\big\{0,0,2\sqrt{2},2\sqrt{2}\big\}$, so
 this indeed represents a physical PPT map.  Combining the channel
 (\ref{eqn:basicexample}) with a 50\% attenuation channel results in a
 channel acting on three-mode covariance matrices as
\begin{equation}
\gamma \mapsto \left(X \oplus \smfrac{1}{\sqrt{2}} \id_2\right)\gamma\left(X \oplus \smfrac{1}{\sqrt{2}} \id_2\right)^t + Y \oplus \smfrac 12 \id_2. \label{eqn:combinedchannel}
 \end{equation}
In the appendix, we show how to derive the action of the complementary
channel using a symplectic representation as in
Figure~\ref{fig:symplectic}.  Then we describe a family of three-mode
covariance matrices that achieve .05 bits of coherent information at
input power $\approx 60$ photons/channel use, and over .06 bits with
input power $\approx 812$ photons/channel use.

%Because $Y + i(J + XJX^t)$ is not full rank, this channel is on the
%boundary of the set of PPT channels and may therefore not be robust
%to experimental errors.  Indeed, recent experimental claims
%\cite{AB09} to having produced the Smolin state, a four-qubit bound
%entangled state discovered by one of us \cite{Smo01}, were subject to
%some debate \cite{LKPR10,AB10} for this very reason.  

\begin{figure}
\hspace{.5in}
\subfloat[]{
\!\!\!\includegraphics[width=2.2in]{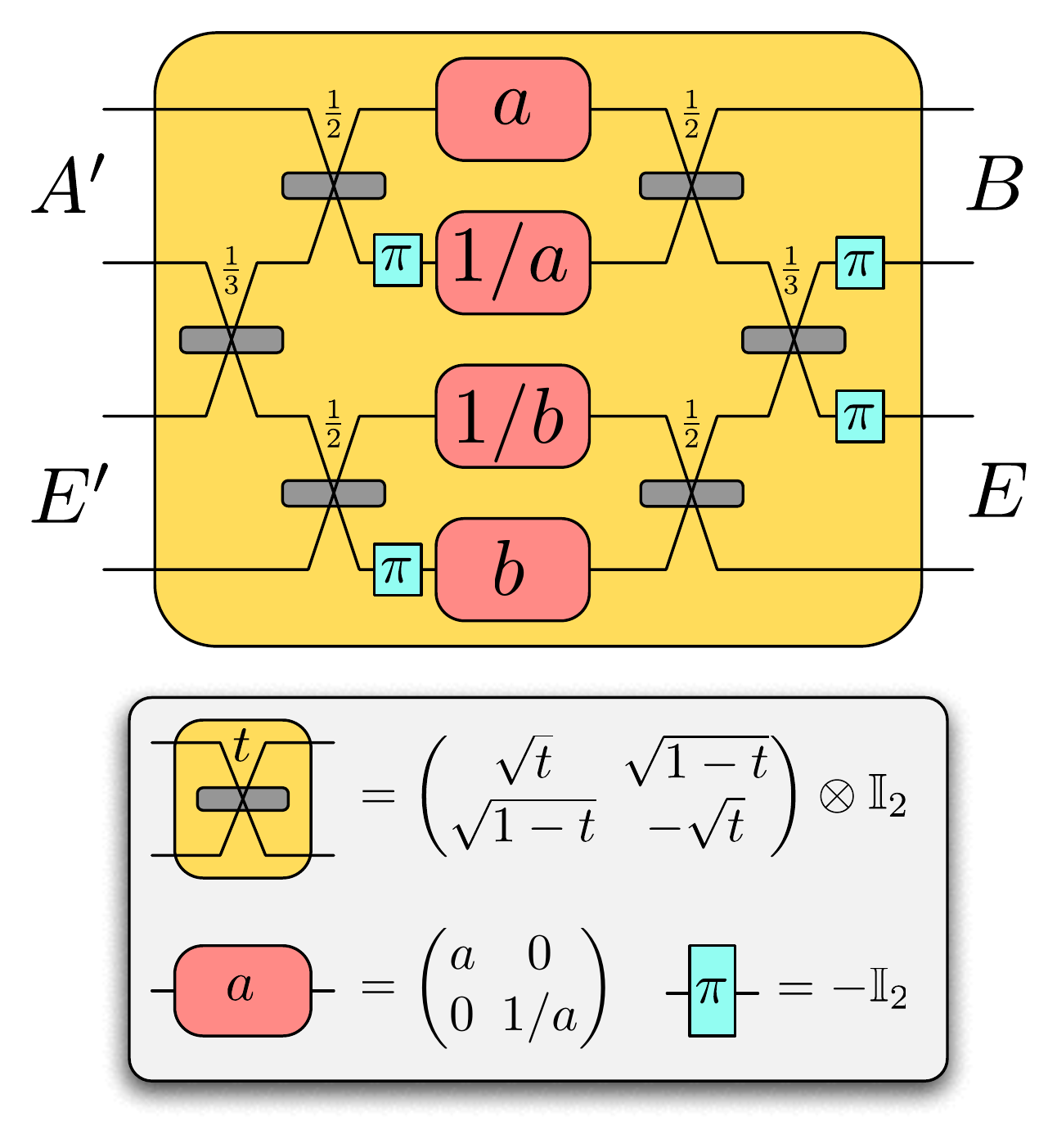}
}
\hspace{.5in}
\subfloat[]{
\includegraphics[width=3.4in]{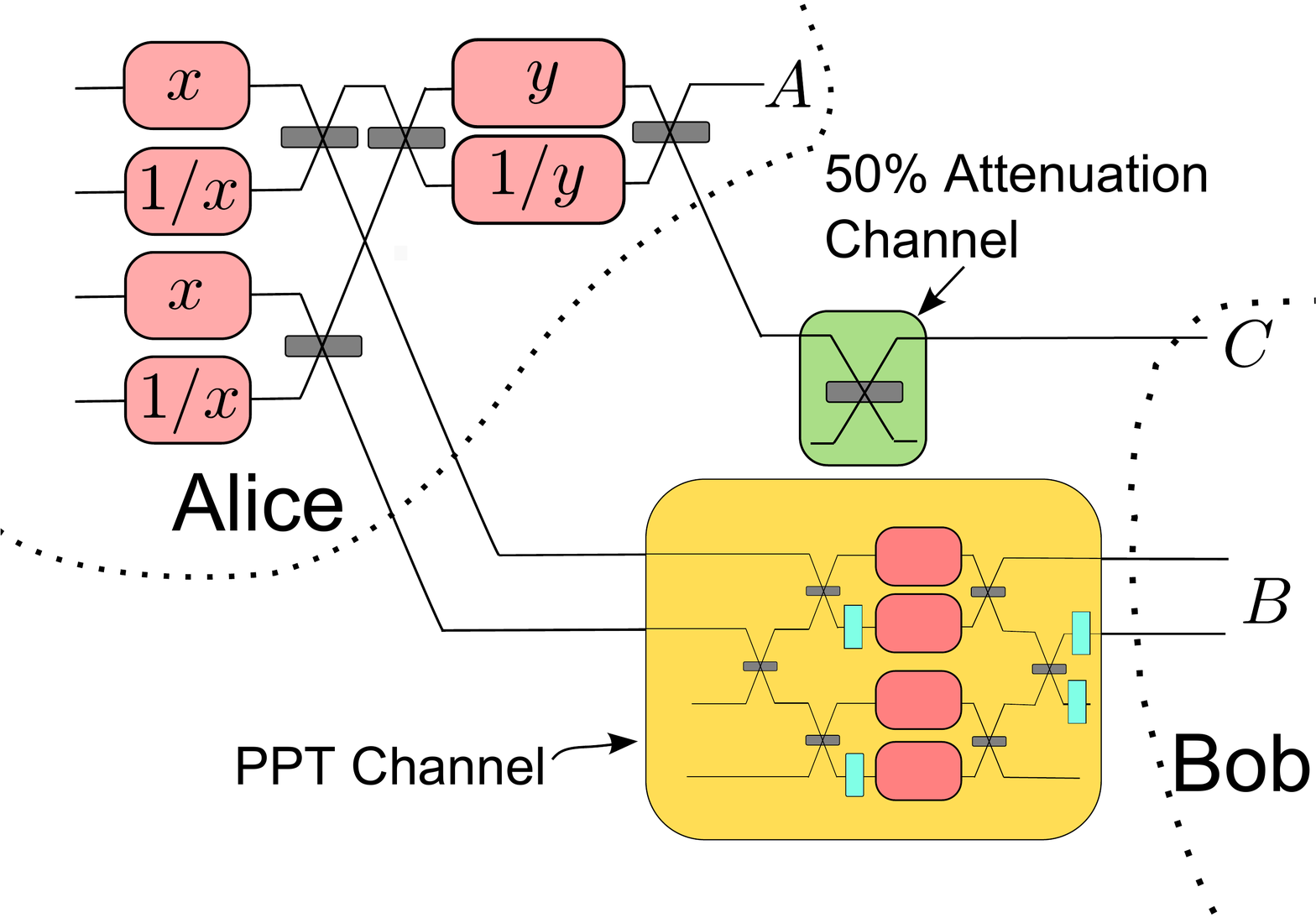}
}
\caption{ \small
A two-parameter family of optical circuits implementing a symplectic transformation in the sense of Figure~\ref{fig:symplectic}. For a range of parameters, we will see that the associated channel is in the interior of the set of PPT channels.   At the bottom, we give the explicit symplectic transformation associated to the three building blocks of our circuits -- transmissivity $1-t$ beamsplitters, nonlinear single-mode squeezing, and half-wave phase plates.  The example (\ref{eqn:basicexample}) is related to the channel with $(a,b) = \left(\sqrt{3} + \sqrt{2},\frac{\sqrt{3}+1}{\sqrt{2}}\right)$ by a canonical transformation. (b)  A family of circuits using the PPT and attenuation channels to generate coherent information between the purification $A$ of the input and the channel outputs $BC$.  Each beamsplitter has parameter $t = \frac 12$.  We see below that superactivation is possible for a flexible range of parameters. }

\label{fig:circuit}
\end{figure}

\begin{figure}[!t]
\centering
\vspace{-.3in}
\includegraphics[width=2.9in]{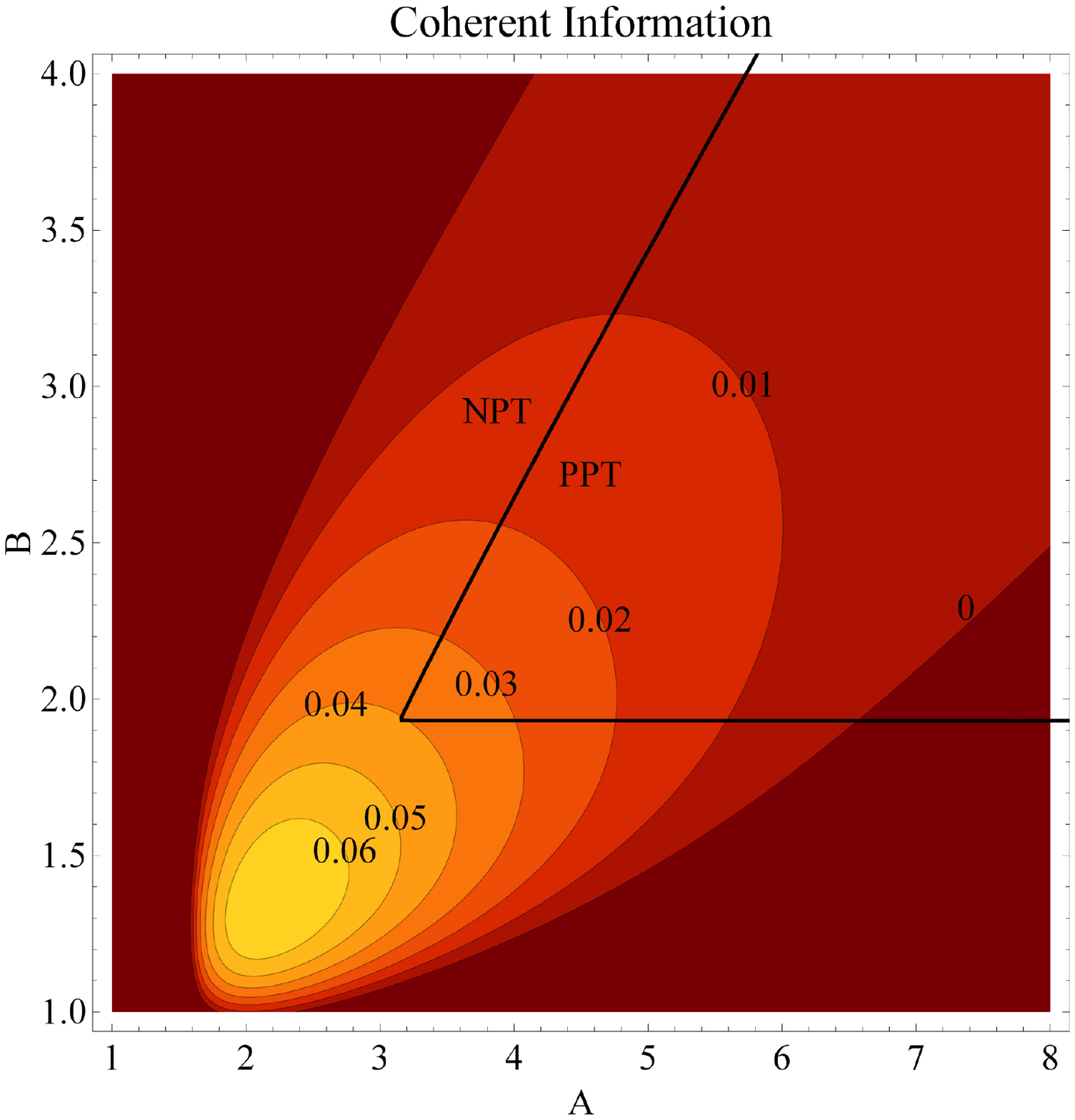}\includegraphics[width=2.9in]{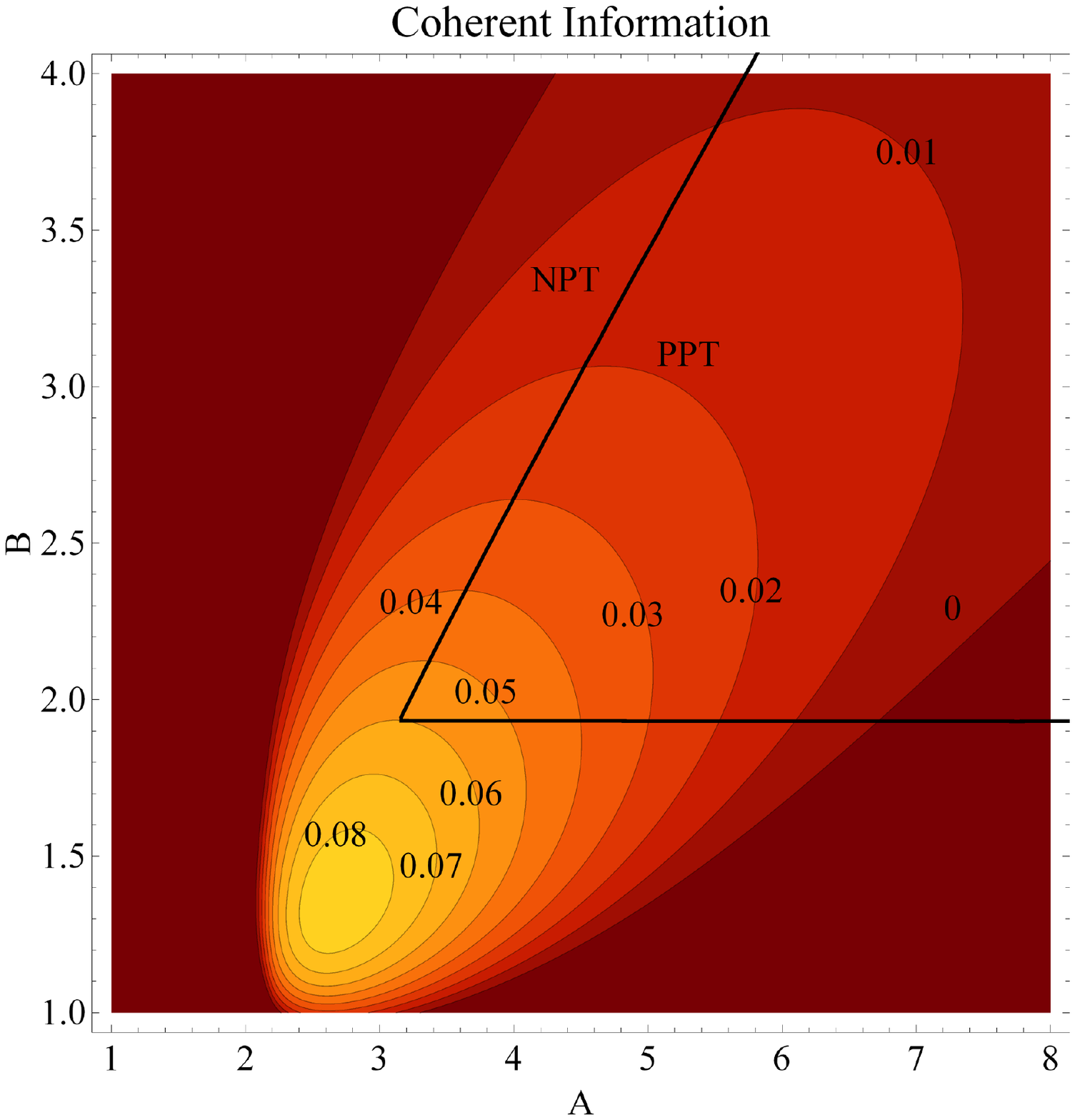}
\vspace{-1.1in}
\caption{\small Superactivation for a wide range of parameters.  Each plot shows the coherent information generated by the family of channels of Figure 2(a) using the strategies of Figure 2(b).
Superactivation occurs in the triangular region to the upper right where the channels are PPT.  The corner of this region is at $(a,b) = \left(\sqrt{3} + \sqrt{2},\frac{\sqrt{3}+1}{\sqrt{2}}\right)$ and corresponds via symplectic transformations to the example (\ref{eqn:basicexample}).  The plot in (a) shows the coherent information for moderate values of squeezing, $x=3$ and $y=3$.
The plot in (b) is similar but for large $x=20$ and optimal value $y = 2 + \sqrt{3}$.}
\label{fig:contourplots}
\end{figure}

Because the matrix $Y +i(J+XJX^t)$ is not full rank, the channel (\ref{eqn:basicexample}) is on the boundary of the PPT channels and any claim of superactivation will be sensitive to experimental errors.  In
Figure~\ref{fig:circuit}, we present a family of symplectic
transformations that include (\ref{eqn:basicexample}) as a special
case and are otherwise experimentally robust.  As any symplectic
transformation can be implemented physically as a combination of
passive linear optical elements (beam splitters and phase shifters)
together with single-mode squeezing, and we present our examples as
circuits of this sort.  
In Figure~\ref{fig:contourplots}, we show the positive coherent information
generated by the examples of Figure~\ref{fig:circuit} for a range of
squeezing parameters and for reasonable input powers, supporting the
notion that that superactivation is indeed generic.  
While linear
optics are straightforward to implement in a laboratory, the nonlinear
squeezings required by our examples will present more of a challenge.
It is nonetheless possible to generate squeezing on the order of
$10$dB with current technology \cite{Valb08}, corresponding to the map $(P,Q) \rightarrow \big(\sqrt{10}P,\frac{1}{\sqrt{10}}Q\big)$, and our examples generally require
squeezing of this order. 
Although an example using only linear optical elements would be desirable, we suspect, but cannot prove, that none exist.

%%%%%%%%%%%%%%
% Figure: circuits
%%%%%%%%%%%%%%

%%%%%%%%%%%%%
% Figure: Contour plots
%%%%%%%%%%%%%

\begin{figure}
\centering
\includegraphics[width=2in]{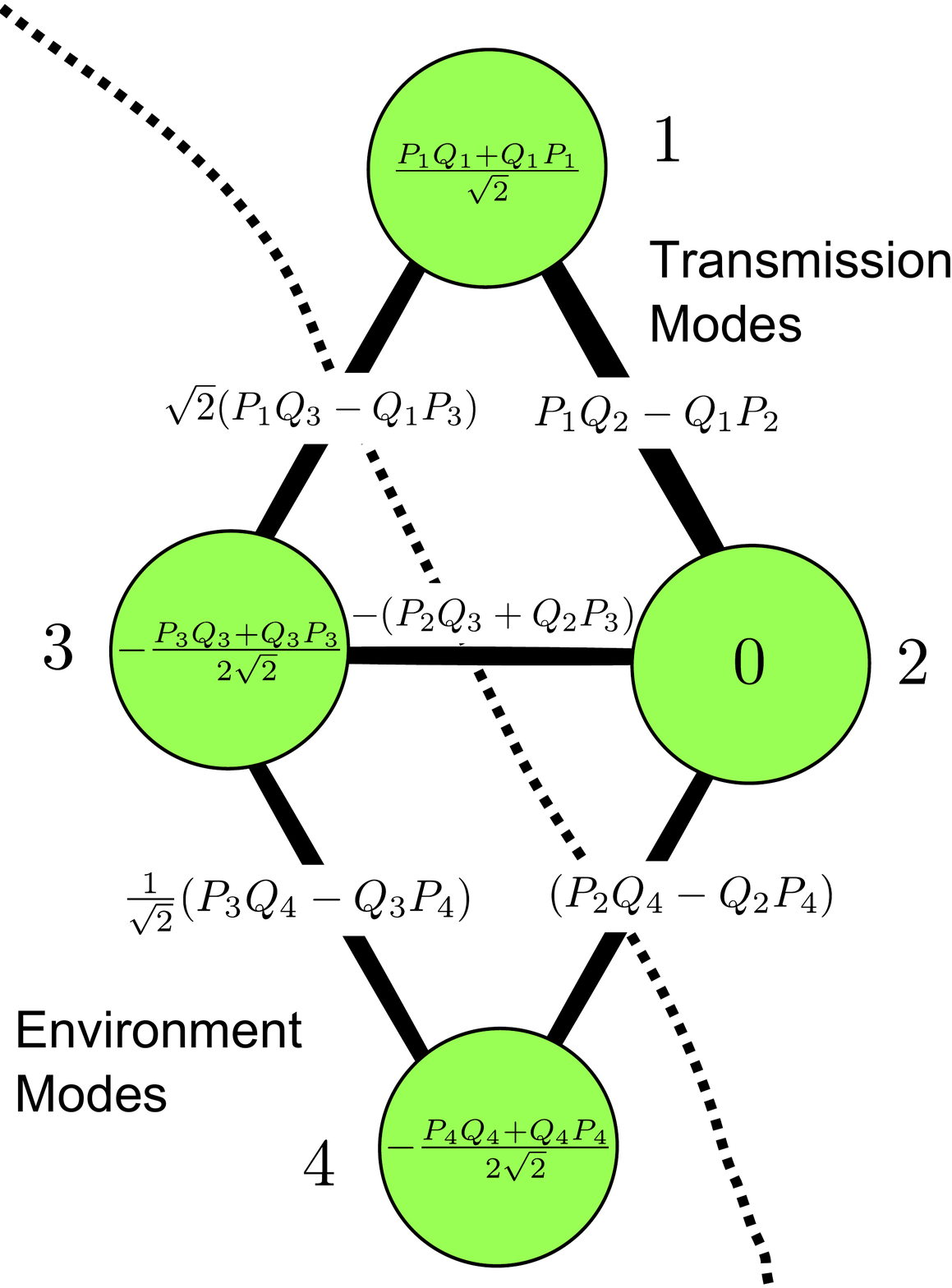}
\caption{\small Generating Hamiltonian.  The full Hamiltonian is the sum of
  the individual and pairwise interaction terms shown.  If environment
  modes 3 and 4 begin in the vacuum state and this interaction runs
  for a time $\pi$, the resulting noisy evolution of transmission
  modes 1 and 2 is given by the $(a,b) = \left(\sqrt{3} + \sqrt{2},\frac{\sqrt{3}+1}{\sqrt{2}}\right)$ channel from Figure \ref{fig:circuit}.}
\label{Fig:Ham}
\end{figure}

We can also analyze how our channels can arise from the continuous
interaction between transmission and environment modes.  This raises
the possibility of our channel occurring ``in the wild.''  Since
quadratic Hamiltonians are ubiquitous, it could be that real-world
optical systems require using superactivation to achieve optimal
performance.  For the purpose of illustration, Figure \ref{Fig:Ham}
shows a natural Hamiltonian that implements the channel in this family
with $(a,b) = \left(\sqrt{3} +
\sqrt{2},\frac{\sqrt{3}+1}{\sqrt{2}}\right)$ after evolution for a
time $\pi$.  

Even before the superactivation result of \cite{SY08}, it was
known that the existence of bound-entangled states with negative
partial transpose (NPT) would imply superactivation of distillable 
entanglement \cite{SST00}.  To date no such NPT quantum states
or channels have been found.  But it is known that there are
no NPT bound entangled gaussian channels \cite{WW01}.  This might have
suggested that the gaussian channels would be too simple for superactivation
occur.  As we have shown, this is not the case.
In \cite{SY08}, superactivation was shown to be a consequence of the
existence of PPT channels with private capacity \cite{HHHO03,HPHH05}.
Rather than pursuing this idea, our approach has been to demonstrate
superactivation directly.  Notably, we don't know whether our channels
have any private capacity or, for that matter, whether there are any
PPT gaussian channels with positive private capacity.  

Our results show that, far from a purely mathematical or singular
phenomenon, superactivation arises naturally for a range of parameters in gaussian 
bosonic systems.  Because it
occurs in systems that seemed to be too noisy to be useful, superactivation
points to the possibility of powerfully enhanced error correction for 
quantum memories and repeaters in the very noisy regime.  It also unveils
an unforeseen complexity in the theory of quantum mechanics with
gaussian states.

\section*{Acknowledgments}
We are grateful to J. Eisert, and M.M. Wolf for helpful advice in the
early stages of this work and C.H. Bennett for many useful
suggestions.  GS and JY are especially thankful to the Institut Mittag Leffler, where some of this work was performed, for their hospitality.  JY's research was supported by grants through the LDRD program of the US Department of Energy.  GS and JAS 
were supported by DARPA QUEST contract HR0011-09-C-0047.

\newpage

\newpage
%%%%%%%%%%%%%%%%%%%%%%
% Appendix
%%%%%%%%%%%%%%%%%%%%%%

\section*{Appendix}
Evaluating the coherent information over the channel (\ref{eqn:basicexample}) on the family of gaussian states with covariance 
 \begin{equation}
 \gamma = 
 \left(
\begin{array}{cccccc}
 7 \cosh (c) & 0 &  \alpha _+\sinh (c) & 0 & \alpha _-\sinh (c)  & 0 \\
 0 & 7 \cosh (c) & 0 & - \alpha _+\sinh (c) & 0 & \alpha _-\sinh (c)  \\
 \alpha _+\sinh (c)  & 0 & \sqrt{2} \cosh (c) & 0 & \cosh (c) & 0 \\
 0 & -\alpha _+\sinh (c)  & 0 & \sqrt{2} \cosh (c) & 0 & -\cosh (c) \\
 \alpha _- \sinh (c) & 0 & \cosh (c) & 0 & \sqrt{2} \cosh (c) & 0 \\
 0 &  \alpha _-\sinh (c) & 0 & -\cosh (c) & 0 & \sqrt{2} \cosh (c)
\end{array}
\right),
\label{eqn:inputstate}
 \end{equation}
where $\alpha_\pm = \sqrt{\frac{7}{\sqrt{2}} + 2\sqrt{3} \pm \frac 12}$,
gives superactivation for all positive input powers, as illustrated here:
\begin{figure}[h]
\centering
\hspace{-1in}
\includegraphics[width=4in]{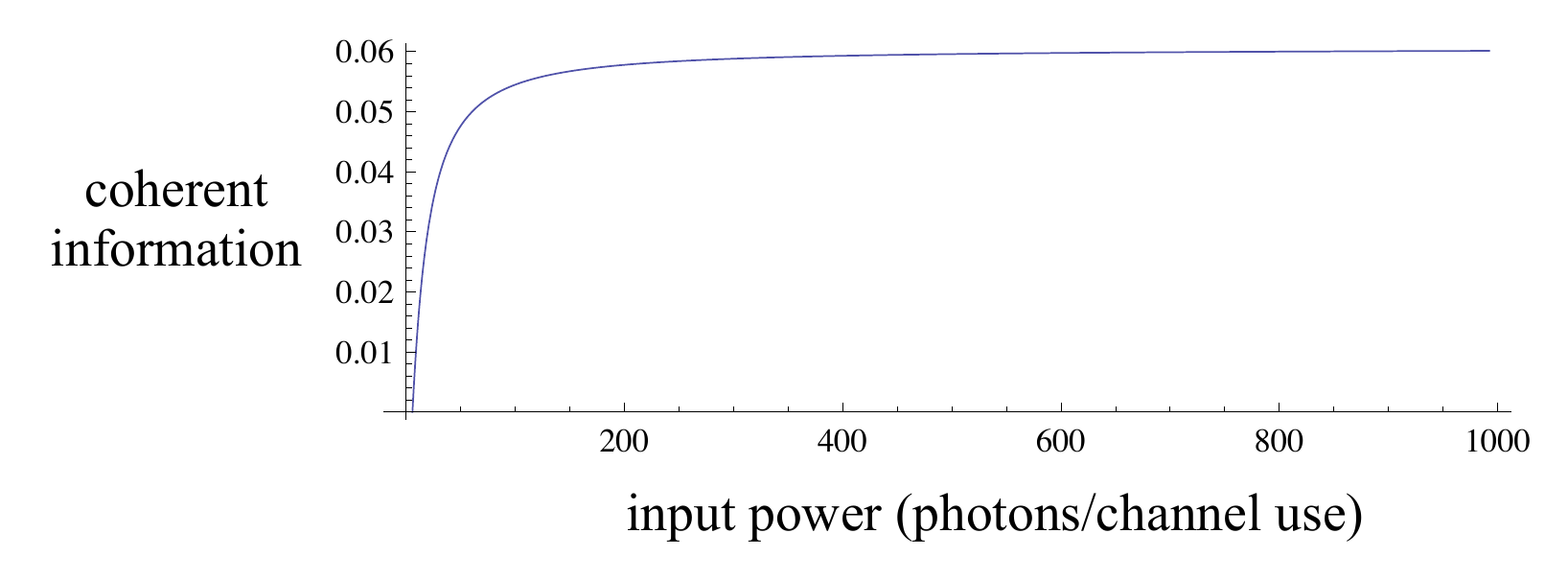}
%\caption{Coherent information over the channel (\ref{eqn:basicexample}) as a function of input power for the family of input states $\gamma$ given in (\ref{eqn:inputstate})}
%\label{fig:powerplot}
\end{figure}

\noindent In particular, this achieves $.05$ bits at $c = 3.19$, or at input power of about 60 photons/channel use, and achieves $.06$ bits at $c=5.8$, or about $800$ photons/channel use.  To compute the coherent information, one also requires an expression for the complementary channel to (\ref{eqn:basicexample}).  This can be obtained from a symplectic extension in the sense of Figure~\ref{fig:symplectic}.  A particularly simple extension exists in this case, given by the block-diagonal symplectic matrix 
\[\pmat{X & Z \\ Z & X}, \,\,\,\,\,\,\,\text{ where }\,\,\,\,\,\,\, Z = 
\pmat{\beta_+  & 0 & \beta_- & 0\\ 
0 & \beta_- & 0 & \beta_+ \\ \beta_- & 0 & -\beta_+
 & 0 \\ 0 & \beta_+ & 0 & -\beta_-},
\]
and $\beta_\pm = \sqrt{ \frac{1}{\sqrt{2}} \pm \frac 12}$, with $X$ as given in (\ref{eqn:basicexample}).

Computation of the coherent information requires computing the von Neumann entropy $H(\rho) = -\Tr \rho \log_2\rho$ of a gaussian state $\rho$.  For a state of $m$ modes, this can be computed starting with the covariance matrix $\gamma$ as follows.  The eigenvalues of $J\gamma$ come in complex conjugate pairs $\pm i \lambda_j$, where the $\lambda_j$ are called the symplectic eigenvalues of $\gamma$.  There then exists a symplectic matrix $S$ such that 
\[S\gamma S^t = \lambda_1 \id_2 \oplus \cdots \oplus \lambda_m \id_2.\]
The von Neumann entropy for such states has the simple form
\[H(\rho) = \sum_j  \left(\smfrac{\lambda_j + 1}{2}\right) \log \left(\smfrac{\lambda_j + 1}{2}\right) - \left(\smfrac{\lambda_j - 1}{2}\right) \log \left(\smfrac{\lambda_j - 1}{2}\right).\]

The channel described in (\ref{eqn:basicexample}) is related to the $(a,b) = \left(\sqrt{3} + \sqrt{2},\frac{\sqrt{3}+1}{\sqrt{2}}\right)$ point in our family of examples as follows.  The latter channel is explicitly described by the matrices 
\begin{equation}\label{eqn:basicexample2}
X' = \left( \begin{matrix} \sqrt{2}& 0 & 1 &0\\ 0 & -\sqrt{2}& 0 & 1\\ -1 &0 & 0 & 0\\ 0 & -1 & 0 & 0 \end{matrix} \right) 
\hspace{1cm} Y' = \left( \begin{matrix}2 &0& -\sqrt{2}& 0 \\ 0& 2 &0& \sqrt{2} \\ -\sqrt{2} &0& 2& 0 \\ 0 &\sqrt{2}& 0& 2 \end{matrix}\right).
\end{equation}
These are related to the matrices (\ref{eqn:basicexample})
by the transformation $X = -S X' S^{-1}T$, $Y = S Y' S^t$, where 
$T$ is the matrix of a $50\%$-beamsplitter and 
\begin{equation}
S = \pmat{\beta_+  & 0 & \beta_- & 0\\ 
0 & \beta_+ & 0 & -\beta_- \\ \beta_- & 0 & \beta_+
 & 0 \\ 0 & -\beta_- & 0 & \beta_+},
\label{eqn:symplecticbasischange}
\end{equation}
is a symplectic transformation corresponding to a two-mode squeezing operation. The matrix $S$ can be viewed as diagonalizing the noise term $Y'$ and is responsible for the the simple form of the example (\ref{eqn:basicexample}).

%\begin{comment}
\begin{figure}
\centering
\includegraphics[width=4in]{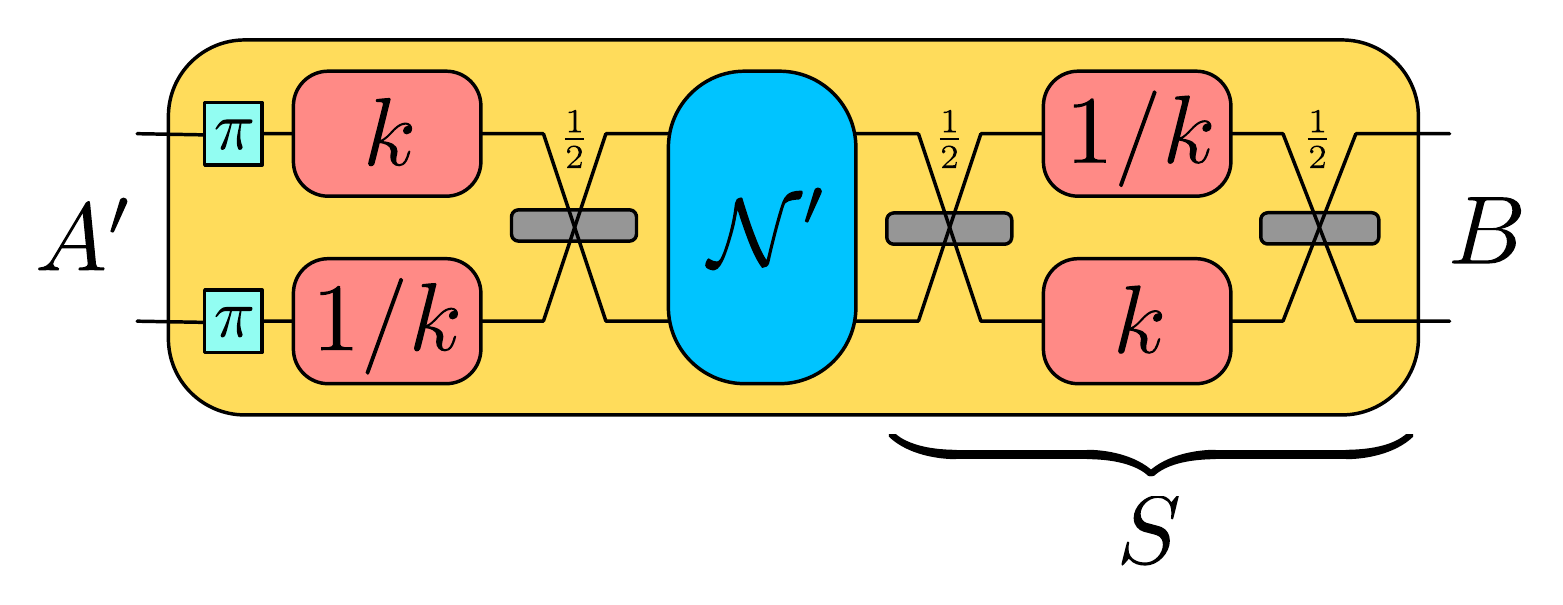}
\caption{\small The channel (\ref{eqn:basicexample}) is related to the channel $\CN'$ at the point $(a,b) = \left(\sqrt{3} + \sqrt{2},\frac{\sqrt{3}+1}{\sqrt{2}}\right)$ in Figure~\ref{fig:circuit} by symplectic transformations at the input and output, pictured above with $k=\sqrt{\sqrt{2}-1}$.  As these correspond to Hilbert space unitaries, the capacity properties of the channels are the same.  The matrix $S$ in (\ref{eqn:symplecticbasischange}) corresponds to the two-mode squeezing primitive at the output of the channel $\CN'$.}
\label{fig:two-mode}
\end{figure}

%\end{comment}

%\bibliographystyle{unsrt}
%\bibliography{GS}

%\clearpage
%\section*{Appendix}
%\subsection{A primer on bosonic gaussian states and channels}

\end{document}